\documentclass[floatfix,nofootinbib,a4paper,11pt]{revtex4-1}  

 \pdfoutput=1
\usepackage{amssymb}
\usepackage[utf8x]{inputenc}
\usepackage[english]{babel}
\usepackage{amsmath}   
\usepackage{amsfonts} 
\usepackage{graphicx} 
\usepackage{float} 
\usepackage{booktabs} 
\usepackage{color} 
\usepackage{rotating} 
\usepackage{units} 
\usepackage{slashed} 
\usepackage{appendix}
\usepackage[text={7in,10in}]{geometry}
\usepackage{epstopdf}
\usepackage{caption}
\usepackage{hyperref} 
\hypersetup{
    colorlinks=true,
    linkcolor=blue,
    citecolor=magenta,      
    urlcolor=green,
}

\newcommand{\bea}{\begin{eqnarray}} 
\newcommand{\eea}{\end{eqnarray}} 
\newcommand{\beq}{\begin{equation}}
\newcommand{\eeq}{\end{equation}} 
\newcommand{\beqa}{\begin{eqnarray}} 
\newcommand{\eeqa}{\end{eqnarray}} 
\newcommand{\bit}{\begin{itemize}} 
\newcommand{\eit}{\end{itemize}}

\newcommand{\met}{\ensuremath{\slashed{E}_T}}

\begin{document}
  
\title{Analysis of the Bounds on Dark Matter Models from Monojet Searches
at the LHC}
 
\author{Swasti Belwal} 
\email[]{swasti@th.physik.uni-bonn.de} 
\affiliation{BCTP, Physics Institute, University of Bonn, Germany } 
   
\author{Manuel Drees} 
\email[]{drees@th.physik.uni-bonn.de} 
\affiliation{BCTP, Physics Institute, University of Bonn, Germany } 
  
\author{Jong Soo Kim} 
\email[]{jongsoo.kim@tu-dortmund.de} 
\affiliation{National Institute for Theoretical Physics,\\
  School of Physics and Mandelstam Institute for Theoretical Physics,\\
University of the Witwatersrand, Johannesburg, South Africa} 

\begin{abstract}

  We analyse the constraints on models of WIMP Dark Matter that can be
  derived from upper bounds on the ``monojet'' cross section at the
  LHC. These constraints were originally interpreted in the context of
  an effective field theory (EFT) where the Standard Model is extended
  by a dimension--6 operator whose coefficient is $1/\Lambda^2$. We
  show that combining the 8 TeV data of the ATLAS and CMS
  collaborations improves the bounds only slightly. We then analyze
  this final state in the context of simplified models with
  $s-$channel mediator. We show that if the decay width of the
  mediator is small, these simplified models can be accurately
  modeled by the effective field theory only if the mediator mass is
  above 5 TeV. Finally, we point out that even if the EFT accurately
  describes the ${\cal O}(\Lambda^{-2})$ contributions to the matrix
  element, for values of $\Lambda$ near the current bound it receives
  significant contributions of order $\Lambda^{-4}$; in the context of
  simplified models, these correspond to diagrams where two mediators
  are exchanged. This observation challenges the internal consistency
  of the EFT description since dimension$-8$ operators, which would
  also contribute to ${\cal O}(\Lambda^{-4})$ to the matrix element,
  are not included.

\end{abstract} 
  
\maketitle 

\section{Introduction} 
\label{sec:intro}

A large number of astrophysical and cosmological observations indicate
that most of the matter in the Universe is non--baryonic
\cite{Drees:Gerbier}. The Standard Model of particle physics (SM) does
not contain a particle that could form this ``Dark Matter''. The
probably most widely studied candidate particle is a Weakly
Interacting Massive Particle (WIMP), with mass in the GeV to TeV range
and with very roughly weak--strength annihilation cross section into
SM particles. The latter implies that, within standard cosmology,
WIMPs that were in thermal equilibrium have approximately the correct
relic density (the so--called ``WIMP miracle'')
\cite{Ade:2013zuv}\cite{Steigman:2012nb}. Moreover, many extensions of
the SM that address some of its other shortcomings automatically
contain, or easily accommodate, viable WIMP candidates \cite{Drees:Gerbier}\cite{Bertone:2004pz}.

Another reason for the popularity of WIMPs as Dark Matter candidates
is that there are several distinct methods to search for them, owing
to their significant interactions with SM particles. In particular,
WIMP interactions with quarks, and hence nucleons and nuclei, allow
Dark Matter WIMPs from the halo of our galaxy to scatter off ordinary
matter, depositing a few (tens of) keV of energy. So--called direct
(WIMP) detection experiments search for this signature
\cite{Drees:Gerbier}. The very same interactions should also allow to
produce pairs of WIMPs\footnote{Single WIMP production is not
  possible as the time reversed version of such matrix elements
  would allow WIMPs to decay, ruling them out as Dark Matter
  candidates.} at hadron colliders like the LHC. Since WIMPs must be
stable on collider time scales and electrically neutral, they escape
detection. The production of a WIMP pair must therefore be tagged by
the emission of a high $p_T$ object recoiling against the WIMP pair,
which manifests itself as missing $p_T$. The largest rates, and hence
the best signals for WIMPs that interact with quarks, result when the
WIMPs recoil against a jet, leading to a monojet signal
\cite{Bai:2010hh}\cite{Beltran:2010ww}.

Ambient Dark Matter WIMPs are non--relativistic, with typical velocity
$v \sim 10^{-3} c$. Hence the maximal momentum exchanged in direct
detection experiments is of order $100$ MeV (for scattering of a
relatively heavy WIMP on a heavy target nucleus like Xenon). Since
WIMPs carry neither electric nor color charge, their interaction with
quarks must be mediated by a massive particle. In most cases the mass
of this mediator is much larger than $100$ MeV. In this case, direct
detection experiments can be analyzed in an effective field theory
(EFT) where the mediators have been integrated out, giving rise to
higher--dimensional operators. This simplifies the analysis since EFTs
often have fewer free parameters than complete models; moreover,
several models may lead to the same EFT, and can thus be treated
simultaneously.

Monojet signals have therefore also been analyzed in terms of an EFT,
both by theorists \cite{Bai:2010hh}\cite{Beltran:2010ww} and by the
LHC collaborations
\cite{Khachatryan:2014rra}\cite{Aad:2015zva}. Clearly this can be
expected to accurately reproduce the results obtained in a
renormalizable theory only if the momentum flow through the mediator
is (much) smaller than the mass of this mediator, as we just saw for
WIMP scattering off baryonic matter. The most sensitive search region
for the 8 TeV data requires missing transverse momentum of about 500
GeV. One may then conclude that the EFT description should work if the
mediator mass is (well) above 500 GeV. In fact, this estimate is not
that far off for $t-$channel mediators
\cite{Papucci:2014iwa}. However, we will see below that the exchange
of narrow $s-$channel mediators can be accurately modeled by an EFT
only for mediator masses above $5$ TeV. In this case the experimental
bounds on $\Lambda$, which are around $1$ TeV, can be saturated only
if the mediator's couplings to quarks and/or WIMPs are considerably
above $1$, i.e. for strongly interacting theories. Moreover, we find
that even in this case some contributions to the matrix element for
WIMP production that are of order $\Lambda^{-4}$ are sizable. This
means that an accurate EFT treatment would need to include
dimension$-8$ operators, thereby introducing (many) new parameters,
spoiling the main advantage of EFTs. Simply ignoring all these
$\Lambda^{-4}$ terms, which seems to have been standard practice in
the experimental analyses, thus means that the EFT as applied to
monojet searches does not accurately describe {\em any} renormalizable
theory with an $s-$channel mediator, if the scale $\Lambda$ is near
the experimental lower bound.

The reminder of this paper is organized as follows. In the next
Section we discuss the theoretical set--up, followed by a description
of the technical details of our analysis in Sec.~3. In Sec.~4 we show
that combining 8 TeV data from the ATLAS and CMS collaborations only
slightly strengthens the lower bound on the EFT scale
$\Lambda$. Sec.~5 contains a comparison of EFT results with results
obtained in simplified models with pseudoscalar or axial vector
$s-$channel mediators. The impact of ${\cal O}(\Lambda^{-4})$ contributions
to the matrix element are discussed in Sec.~6. Finally, we summarize our
results and draw some conclusions in Sec.~7.

\section{Theoretical Framework}
\label{sec:framework} 

We are interested in theories of WIMP Dark Matter interacting with
quarks. Such theories can be tested both via direct WIMP searches, and
via searches for events with missing transverse momentum at hadron
colliders such as the LHC. For definiteness we assume that the WIMP is
a Dirac fermion $\chi$. Models with Majorana fermions would lead to
very similar results for the interactions we consider. Models with
(real or complex) scalar WIMP obviously do not lead to spin--dependent
contributions to the matrix elements for WIMP--nucleon scattering;
such WIMPs cannot interact with protons via the exchange of
pseudoscalar or axial vector mediators. UV-complete models where the
WIMPs carry spin of one unit are a bit more complicated, without
adding anything fundamentally new to our study.

An effective field theory (EFT) describing the interactions of $\chi$ with standard
quarks $q$ can then be written as \cite{Bai:2010hh} :
\begin{equation} \label{eq:eft}
{\cal L}_{\rm EFT} = \sum_{\Gamma} \frac {1} {\Lambda_\Gamma^2} \bar \chi
\Gamma \chi \bar q \Gamma q\,.
\end{equation}
Here
$\Gamma \in \{1, i \gamma_5, \gamma_\mu, \gamma_\mu \gamma_5,
\sigma_{\mu\nu}\}$
for scalar, pseudoscalar, vector, axial vector, and tensor
interactions, respectively. In principle WIMPs could couple to
different quarks with different strengths, i.e. the parameters
$\Lambda_{\Gamma}$ could depend on the flavor of $q$ as well; however,
we will assume that all quarks couple with equal strength, or not at
all, to the WIMPs [i.e. some heavy quarks may not appear in the
effective Lagrangian of eq.(\ref{eq:eft})]. Moreover, in order to
further simplify the analysis, we will assume that only one of the
operators in eq.(\ref{eq:eft}) is present, i.e. all but one of the
$\Lambda_\Gamma$ will be sent to infinity.

The effective Lagrangian of eq.(\ref{eq:eft}) allows to describe the
production of a $\chi \bar \chi$ pair in $q \bar q$ annihilation. If
$\chi$ is to be a Dark Matter particle, it should be electrically and
color neutral, and stable on collider time scales. Such a particle
will traverse an LHC detector without trace. Inclusive
$\chi \bar \chi$ pair production thus looks like producing
``nothing'', as far as the LHC detectors are concerned. Clearly this
is not a viable signature.

\begin{figure}	
\begin{center}
	\includegraphics[scale=0.55]{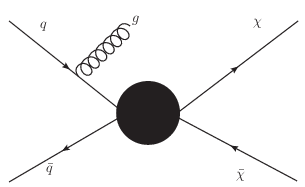}
	\caption{A Feynman diagram giving rise to a monojet event when
          the WIMPs ($\chi, \bar \chi$) escape detection. The blob
          stands for one of the four--point EFT interaction vertices of
          eq.(\ref{eq:eft}).}
\label{fig:eft_vertex}
\end{center}
\end{figure}

The signature becomes viable if ``nothing'' recoils against a
high$-p_T$ object.  The largest cross section, and strongest bound,
results when this object is a jet, which can result from the emission
of a single high$-p_T$ parton.  Since the WIMPs escape detection, this
leads to the celebrated ``monojet'' signature, where the event
contains a single hard jet, leading to a large amount of missing
transverse momentum. A Feynman diagram leading to this signature in
the framework of the EFT of eq.(\ref{eq:eft}) is shown in
Fig.~\ref{fig:eft_vertex}.

One of the goals of our analysis is to compare predictions derived
from the EFT defined by eq.(\ref{eq:eft}) with those derived from a
``simplified model'', where $\chi$ interacts with quarks via the
exchange of one mediator. Here we focus on $s-$channel mediators for
two reasons. First, $t-$channel mediators would need to carry color,
whereas $s-$channel mediators do not. The latter can for e.g. be
additional gauge or Higgs bosons, which have been widely discussed in
the literature for reasons not related to Dark Matter; such models
therefore seem somewhat better motivated than models with $t-$channel
mediators.\footnote{An exception here is supersymmetry which
  automatically contains $t-$channel mediators, namely the
  squarks. However, even the simplest potentially realistic
  supersymmetric extension of the SM, the minimal supersymmetric
  standard model or MSSM \cite{Csaki:1996ks}, also contains
  $s-$channel mediators, since the WIMPs also couple to $Z$ and
  neutral Higgs bosons. The MSSM can therefore in general not be
  described by a simplified model.} Secondly, in models with
$t-$channel mediators a Fierz transformation is required to bring the
effective Lagrangian into the form of eq.(\ref{eq:eft}), which has
also been used by the experimental groups.  Hence a simplified model
with a $t-$channel mediator will generally produce several terms in
the effective Lagrangian simultaneously, thereby complicating the
analysis.

Among the $s-$channel mediators, models with a scalar (CP-even) or
vector mediator will lead to spin-independent contributions to the
WIMP-nucleon scattering matrix elements. There are very strong
constraints on such interactions from direct Dark Matter search
experiments
\cite{Drees:Gerbier}\cite{Liem:2016xpm}\cite{Cheung:2012gi}. Moreover,
there is no renormalizable theory with mediators of spin-2, which
would lead to tensor interactions. Hence we focus on pseudoscalar
(CP-odd) and axial vector mediators. In the subsequent subsections we
discuss these two cases in turn.

\subsection{Axial Vector Mediator} 
\label{subsec:axial-vector_mediator}

The general Lagrangian describing the WIMP $\chi$ interacting with
quarks $q$ via exchange of an axial vector (AV) mediator $A$ is:
\beq \label{eq1}
{\cal L}_{AV} = {\mathcal{L}}_{SM} - \frac{1}{4} A^{\mu\nu} A_{\mu\nu} 
+ \frac{1}{2} M_A^2 A^{\mu} A_{\mu} + \dot\imath \bar{\chi} \gamma^{\mu} 
\partial_{\mu} \chi - m_{\chi} \bar{\chi}\chi + 
g_{\chi A} A_{\mu} \bar{\chi}  \gamma^{\mu} \gamma^5 \chi 
+ \sum_q g_{qA} A_{\mu} \bar{q} \gamma^{\mu} \gamma^5 q\,.
\eeq
Here, ${\cal L}_{SM}$ stands for the general standard model
Lagrangian, $A^{\mu\nu} = \partial^\mu A^\nu - \partial^\nu A^\mu$ is
the field strength tensor, and $g_{qA}$ and $g_{\chi A}$ are the
couplings of the mediator to $q$ and $\chi$, respectively; finally,
$M_A$ and $m_{\chi}$ are the masses of the mediator and the WIMP,
respectively.

We neglect couplings of the mediator to the leptons since they play no
direct role in the monojet signature we will investigate. Moreover,
there are stringent constraints on such couplings from di--lepton
resonance searches \cite{Altmannshofer:2014cla}. As noted above, we
take equal couplings of the mediator to all the quark flavors. When we
integrate out the $A$ field, the effective dimension-6 four fermion
operator corresponding to the Lagrangian of eq.(\ref{eq1}) is
\beq \label{eq2}
{\mathcal{O}}^6_{AV} = \frac{1}{\Lambda^2} \left( \bar{q} \gamma_\mu\gamma_5 
q\right) \left(\bar{\chi} \gamma^\mu \gamma_5 \chi\right )\,.
\eeq
For a weakly coupled theory, the total decay width $\Gamma_A$ of the mediator
should be much smaller than its mass. 
The cut--off scale $\Lambda$ of the effective theory is then
\beq \label{eq3} 
\Lambda  = M_A/\sqrt{g_{\chi A} g_{qA}}\,.
\eeq

If $\Gamma_A$ is not negligible, the numerator in eq.(\ref{eq3})
should be replaced by $\left[ M_A^2 \left( M_A^2 + \Gamma_A^2 \right) 
\right]^{1/4}$. We will see later that this can cause problems if 
$M_A$ is significantly larger than $\Lambda$.

\subsection{Pseudoscalar Mediator} 
\label{subsec:pseudo-scalar_mediator}

The case of a real spin--0 CP--odd pseudoscalar (PS) mediator $P$ is
described by the following Lagrangian:
\beq \label{eq4}
{\cal L}_{PS} = {\cal L}_{SM} + \frac{1}{2} (\partial_{\mu} P)^2 
- \frac{1}{2} M_P^2 P^2 + \dot\imath \bar{\chi} \slashed\partial_{\mu} \chi
- m_{\chi} \bar{\chi} \chi - \dot\imath g_{\chi P} P \bar{\chi} \gamma^5 \chi 
- \sum_{q} \dot\imath g_{qP} P \bar{q} \gamma^5 q\,.
\eeq
Here, $g_{qP}$ and $g_{\chi P}$ are the couplings of the quarks and
the WIMP to the mediator, and $M_P$ and $m_{\chi}$ are the masses of P
and $\chi$, respectively. We again ignore possible couplings of the
mediator to leptons. 

Note that the couplings in eq.(\ref{eq4}) violate chirality. It is
usually assumed that chirality breaking in the SM is governed by the
fermion masses. We therefore follow the usual assumption
\cite{Bai:2010hh}\cite{Aad:2015zva} by setting the couplings of $P$ to
quarks be proportional to the quark masses:
\beq \label{eq4a}
 g_{qP} = g \frac{y_q}{\sqrt{2}}
\eeq
where $y_q$ is the SM Yukawa coupling of quark $q$ and $g$ is a real constant.
Integrating out the $P$ field gives the dimension$-6$ effective operator:
\beq \label{eq5}
{\mathcal{O}}^6_{PS} = \frac{m_q}{\Lambda^3} \left(\bar{q}\gamma_5 q\right) 
\left(\bar{\chi} \gamma_5 \chi\right)\,.
\eeq
The cut--off scale $\Lambda$, which from eq.(\ref{eq4a}) is common for
all quarks (with non--vanishing coupling to $P$) is given by:
\beq \label{eq6}
{\Lambda^3} = \frac{M_P^2 m_q}{g_{qP} g_{\chi P}} = \frac {v M_P^2} 
{g g_{\chi P}}\,,  
\eeq
where $v \simeq 246$ GeV is the vacuum expectation value of the Higgs field
breaking the electroweak gauge symmetry. Here we have again assumed that
the theory is weakly coupled, i.e. that the total decay width $\Gamma_P$ of 
the mediator is significantly smaller than its mass, $\Gamma_P^2 \ll M_P^2$.

It should be noted that simplified models with $s-$channel mediators
necessarily also generate dimension$-6$ operators of type
$(\bar q q) (\bar q' q')$ (where $q'$ and $q$ may be different or
equal quarks) and $(\bar \chi \chi) (\bar \chi \chi)$, with
coefficients that are proportional to the square of the mediator's
coupling to quarks and WIMPs, respectively. To leading order in an
expansion in powers of $\Lambda^{-2}$ these additional operators play
no role in the cross section for monojet production; we will
nevertheless see below that their contribution can be significant for
values of $\Lambda$ near the current lower bound.

\section{Analysis Framework} 
\label{sec:analysis}

In this Section we describe technical details of our analysis of monojet
signals predicted by the models we described in the previous Section.

We begin with writing the model files in FeynRules-v2.0
\cite{Alloul:2013bka}. All the necessary model details for the
particles and their couplings, along with the complete Lagrangian, are
given as an input to FeynRules-v2.0. The output from FeynRules,
describing all the possible interactions in the Universal Feynrules
Output format (UFO) \cite{Degrande:2011ua}, is then used as input to
MadGraph5-aMC.at.NLO.v2.3.0 \cite{Alwall:2014hca} for matrix element
calculations and event generation. The calculation is done at
tree--level. We use the MSTW2008LO set of parton densities
\cite{Martin:2009iq} as implemented in the LHAPDF package
\cite{Andersen:2014efa}.

The monojet signal requires the existence of at least one hard jet,
but events with at least one additional jet are also accepted. We
therefore ask MadGraph to generate events with a $\chi \bar \chi$ pair
plus one or two hard partons in the final state. These events are
passed to PYTHIAv6.4 \cite{Sjostrand:2006za}, which is integrated with
MadGraph5-aMC.at.NLO.v2.3.0, for showering. Hard parton showering off
the $\chi \bar \chi$ plus one parton sample of MadGraph events leads
to the same final state as $\chi \bar \chi$ plus two parton MadGraph
events with only relatively soft showering. In order to remedy this
double counting we use the MLM matching prescription
\cite{Alwall:2007fs} for jets.

Our cross section will diverge as the $p_T$ of the jets goes to zero,
as do all cross sections involving jets at hadron colliders. Since the
experimental monojet searches use strong cuts on the missing $E_T$
($\met$) and on the $p_T$ of the hardest jets, we apply parton--level
cuts at MadGraph level of $p_T$ of 200 GeV for the leading jet and
minimum $\met$ of 300 GeV. We use a value of $100$ GeV for the
\textit{xqcut} variable, which separates the region of phase space to
be populated by showering from that populated by the second hard
parton explicitly generated by MadGraph; we checked that varying the
value of \textit{xqcut} by up to a factor of two has little impact on
the final cross section after all cuts. All other parameters are kept
at their default values. Since the rather stiff generator--level cuts
ensure that the efficiency for passing the final cuts is not very low,
we found it sufficient to generate $50,000$ events per point in
parameter space.

The final selection cuts for the various monojet signal regions as
well as a (simplified) simulation of detector effects are performed
with CheckMATE-v2.0 \cite{Drees:2013wra,Dercks:2016npn} based on
Delphes-v3.0.10 \cite{deFavereau:2013fsa} with modified detector cards
as well as Fastjet-v3.0.6 \cite{Cacciari:2011ma} for the jet
reconstruction. To that end, we implemented the ATLAS
\cite{Aad:2015zva} and CMS \cite{Khachatryan:2014rra} mono--jet
analyses for $\sqrt{s}$=8 TeV in CheckMATE-v2.0 following the
prescription of \cite{Kim:2015wza}. Briefly, both ATLAS and CMS veto
the presence of identified charged leptons, in order to suppress
backgrounds from $W+$jets. Several signal regions are defined, which
differ by the lower bound on the missing $E_T$; the most sensitive
signal regions, which are used to set the final cuts, typically have
$\met > 400$ or $500$ GeV.

It is important to note that neither ATLAS nor CMS strictly speaking
search for pure monojet events. Vetoing events with a second
reconstructed jet would reduce the signal considerably, since many
events with $\met \sim 500$ GeV produce at least one additional jet from
showering. CMS rejects events that have more than two jets with $p_T > 30$
GeV; moreover, if there is a second jet, the azimuthal angle between the
two jets has to be less than $2.5$ radians (or $143$ degrees). ATLAS does not
impose an explicit upper bound on the number of jets but requires that the
$p_T$ of the hardest jet is at least $50\%$ of the total $\met$; moreover,
all jets must have azimuthal angle relative to the missing $p_T$ vector of
at least $1$ radian ($57$ degrees).

\begin{figure}
\includegraphics[scale=0.5]{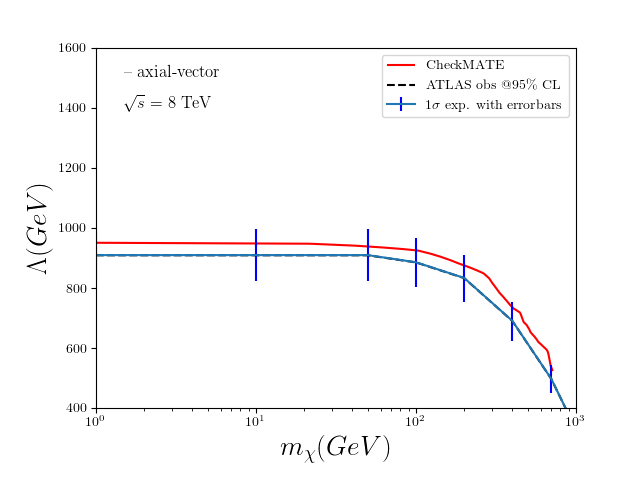}
\includegraphics[scale=0.5]{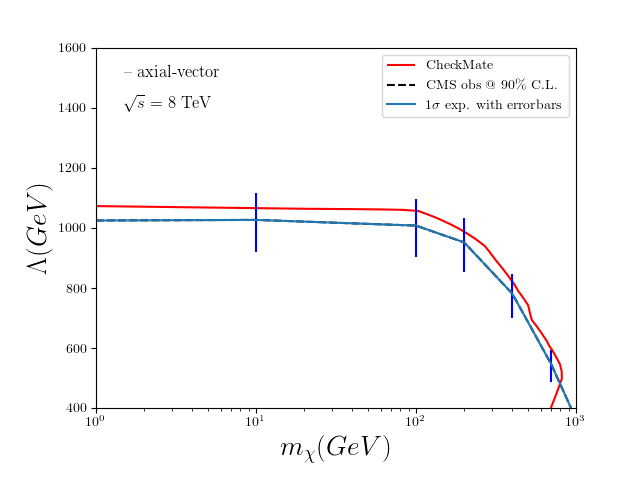}
\caption{Our limits on the strength of the axial vector interaction
  derived using CheckMATE compared to the ATLAS limits on $\Lambda$ at
  $95\%$ CL and CMS limits on $\Lambda$ at $90\%$ CL for $\sqrt{s}=8$
  TeV.}
\label{fig:8TeV}
\includegraphics[scale=0.5]{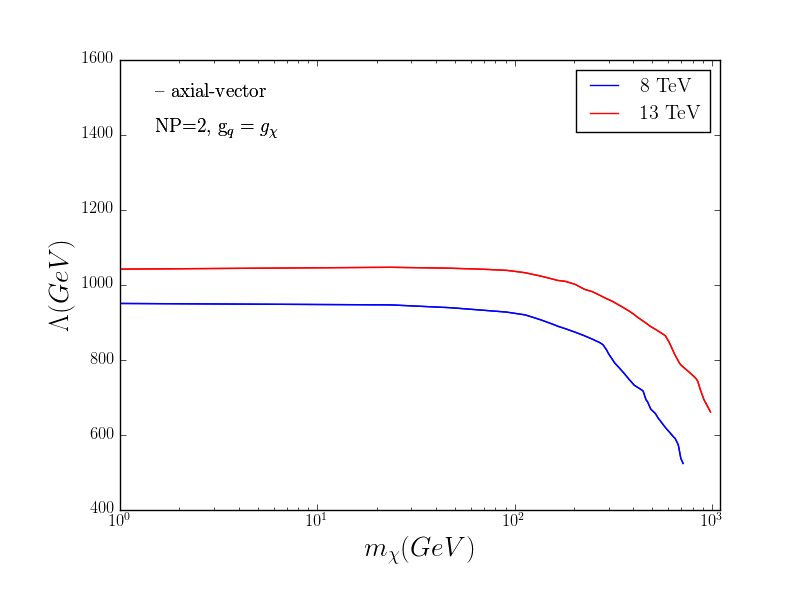}
\caption{Bounds on $\Lambda$ for axial vector interaction at 95\% CL using
ATLAS data taken in 2015 at $\sqrt{s}=13$ TeV.}
\label{fig:13TeV}
\end{figure}

We use the model with axial vector mediator to validate our analysis
chain by comparing our bounds on the scale $\Lambda$ with the
bounds derived by the ATLAS and CMS collaborations, for various values
of the WIMP mass between $1$~GeV and $1$~TeV. Since we wish to
reproduce the EFT limit, we set the mediator mass to $10$~TeV, and its
width to $1$~GeV such that eq.(\ref{eq3}) is applicable. For the
generation of MadGraph events, we set $N_{NP}=2$, i.e. we allow only a
single exchange of the mediator in the signal process. Results for the
$8$ TeV analyses are depicted in Fig.~\ref{fig:8TeV}. We see that we
reproduce the experimental lower bound on the scale to better than
$10\%$. From this figure the CMS limit looks more stringent than that
of ATLAS. Note, however, that ATLAS quotes a lower bound at $95\%$
confidence level (CL), whereas CMS only requires a CL of $90\%$.
 
For the early (2015) $\sqrt{s} = 13$ TeV data, ATLAS and CMS do not
use the EFT interpretation of their monojet bounds any more. We
instead consider the simplified model analysis of ATLAS
\cite{Aaboud:2016tnv}. We implemented their signal regions into
CheckMATE-v2.0 and tested our effective model against this
search. Although at this increased center of mass energy the on--shell
production of the mediator is in principle possible, we found that
this contribution is negligible because of the very small parton
densities at the required large parton energies. Fig.~\ref{fig:13TeV}
shows that the resulting lower bound on the cut--off scale is
approximately $100$ GeV stronger than that derived from data taken at
$\sqrt{s} = 8$ TeV, although the integrated luminosity of the latter
data set is about six times larger.

Owing to the factor of $m_q$ in the coefficient of the pseudoscalar
four fermion operator, see eq.(\ref{eq5}), the resulting bound on
$\Lambda$ is much weaker than in the axial vector case. For example,
even for light WIMP ATLAS quotes a lower bound of just $40$~GeV on
$\Lambda$. This is well below the values of even the basic
acceptance cuts defining monojet events, making the usefulness of an
EFT description in this case a priori unlikely. However, formally our
set--up can also treat this case, by simultaneously choosing a large
pseudoscalar mediator mass, a small mediator width, and very large
values for the coefficient $g$ defined in eq.(\ref{eq4a}).

\section{Combined Analysis} 
\label{sec:a_combined_analysis:}

Since ATLAS and CMS give comparable bounds on the scale $\Lambda$, one
expects to obtain stronger limits by combining both data sets.  Both
ATLAS and CMS use the ``confidence level'' method to set the limits
\cite{Junk:1999kv}\cite{Lista:2016chp}. If the errors on the
background estimates are not correlated between the experiments it is
straightforward to combine the results. The combined total background
error will then be the sum in quadrature of the two separate
errors. This is probably not a bad approximation since the background
estimates are data driven. We use Gaussian statistics for this
computation.

To explain how we estimate the combined limits we briefly describe an
example of combining ATLAS SR7 ($\met > 500$ GeV) \cite{Aad:2015zva}
with CMS SR7 ($\met > 550$ GeV) \cite{Khachatryan:2014rra}. ATLAS
quotes an expected background of $1,030 \pm 60$ events in this signal
region, and finds $1,028$ events. From this we compute an upper bound
on the number of signal events at $95\%$ CL,
$N_{95,{\rm ATLAS}} = 134$ events, to be compared with an upper bound
of $146$ events quoted by ATLAS. CMS expects $509 \pm 66$ events in
their SR7, and observes $519$ events. Our computed
$N_{95,{\rm CMS}} = 145$ whereas CMS cites a value of $142$.

Combining these SRs we have a total expected background of
$1,539 \pm 89$ events, whereas the actual number of observed events is
$1,547$. From this we compute a combined $95\%$ CL upper bound on the
number of signal events $N_{95,{\rm combined}} = 198$, to be compared
with the sum of the individual $N_{95}$ values of $279$ events. The proper
statistical combination thus reduces the upper bound on the total number of
signal events by about $30\%$.

In practice, we let CheckMATE select the two signal regions which are
expected to have the best sensitivity, based on the expected number of
background events. We combine only these two SRs which are
statistically independent since they refer to different
experiments. In this way we avoid ``look elsewhere'' problems that
could arise if we combined all nine ATLAS signal regions with all
seven CMS signal regions. The most sensitive signal region depends on
the value of $m_\chi$, which (for large mediator mass) basically fixes
the kinematics of the process.

Unfortunately, we find that the combination strengthens the bound on
$\Lambda$ in the model with axial vector mediator only very
slightly. For example, for light WIMPs, $m_\chi = 1$ GeV, and again
only including contributions with $N_{NP}=2$, we find a combined
$95\%$ CL lower bound $\Lambda > 970$~GeV, compared to individual
$95\%$ CL lower bounds of $950$~GeV for ATLAS and $900$~GeV for
CMS. This improvement is not really significant.

\section{Comparison Between the EFT and Simplified Models}
\label{sec:a simplified model case}

In this Section we first discuss under what circumstances our simplified
models can be accurately described by an EFT as far as monojet production is
concerned, taking the finite width of the mediator into account. We then
briefly discuss other limits on the models, which have nothing to do with
monojet searches.

\subsection{Finite Width Effects and Applicability of the EFT}

So far we have simultaneously chosen large masses and small widths for
our $s-$channel mediators. This allows to reproduce the EFT limit in
our formalism; note that FeynRules does not allow to directly input
four--fermion operators into the Lagrangian. We also saw that the
current LHC bound on the scale $\Lambda$ is about a TeV for the axial
vector mediator, and only about 40 GeV for the pseudoscalar mediator,
even for light WIMPs; for heavier WIMPs the bounds become even weaker.

However, as pointed out at the end of Sec.~II, requiring the mediator
mass to be significantly larger than $\Lambda$ requires couplings
which are larger than $1$. This in turn leads to large widths of the
mediator. In other words, the combination of a large mediator mass
$M^2 \gg \Lambda^2$ with a small mediator width $\Gamma^2 \ll M^2$ cannot be
realized in a physical model.

In a more realistic situation, i.e. in a real `simplified model', the
width of the mediator is instead a derived quantity
\cite{Harris:2014hga}\cite{Buckley:2014fba}\cite{Abdallah:2014hon}. For
this discussion we consider Dirac fermionic $\chi$ of mass 1 GeV; as
noted above, for heavier $\chi$ the bound on $\Lambda$ is weaker,
making the problem even more severe. This means that the mediator can
always decay into WIMPs as well as into quarks.\footnote{If the decay
  into WIMPs were not possible, one could generate a small width of
  the mediator by choosing its coupling to quarks to be very small. In
  order to keep $\Lambda$ fixed, the coupling strength to WIMPs would
  have to be increased such that the product of the couplings is
  constant. This quickly would require couplings to the WIMP exceeding
  $\sqrt{4\pi}$, again indicating that at least one sector of the
  model is not perturbative.} At tree level, the widths of the
pseudoscalar and axial vector mediators are given by:
\beq \label{eq12}
\Gamma_P = \frac{M_P}{8\pi} \bigg[ g_{\chi P}^2 \bigg( 1 
- \frac{4m_\chi^2}{M_P^2} \bigg)^{1/2} + N_c g^2 \sum_q  
\frac{m_q^2}{v^2} \bigg(1 - \frac{4m_q^2}{M_{P}^2} \bigg)^{1/2}
 \bigg]\, ;
\eeq
\beq \label{eq13}
\Gamma_A = \frac{M_A}{12\pi} \bigg[ g_{\chi A}^2 \bigg (1 
- \frac{4m_\chi^2}{M_A^2} \bigg)^{3/2} + N_c g_{qA}^2 \sum_q 
\bigg( 1 - \frac{4m_q^2}{M_A^2} \bigg)^{3/2} \bigg] \,.
\eeq
We have again assumed that the axial vector mediator has common coupling
$g_{qA}$ to all quarks, whereas the coupling of the pseudoscalar mediator
to quarks is given by eq.(\ref{eq4a}), with $y_q/\sqrt{2} = m_q/v$. The factor
$N_c = 3$ accounts for the color of quarks.

Evidently the decay width of the mediator scales like the squared
coupling times the mass of the mediator. As long as the mediator is
narrow, $\Gamma^2 \ll M^2$, increasing $M$ for fixed $\Lambda$ implies
that the couplings grow proportional to $M$, see eqs.(\ref{eq3})
and (\ref{eq6}). In that case the mediator's width will scale like
$M^3$. This means that the $M^2 \Gamma^2$ term in the squared
propagator of the mediator, as obtained from eqs.(\ref{eq12}) and
(\ref{eq13}), will scale like $M^8$ with increasing mass. When the
mediator's width becomes comparable to its mass a perturbative
treatment is no longer possible; moreover, eqs.(\ref{eq3}) and
(\ref{eq6}), which ignore the $\Gamma^2 M^2$ term in the squared
propagator, are no longer valid.

In order to illustrate this problem, consider the AV case with
$M_A = 10$~TeV, $\Lambda = 1$~TeV and $g_{\chi A} = g_{q A}$ for
all six quark flavors. Eq.(\ref{eq3}) then gives
$g_{\chi A} = g_{q A} = \sqrt{10}$, which via eq.(\ref{eq13}) leads to
$\Gamma_A \simeq 50\ {\rm TeV}\, = 5 M_A$! This is clearly beyond the
domain of perturbation theory, and beyond the domain of applicability
of eq.(\ref{eq3}).

In our earlier analyses we chose $M_A = 10$ TeV just to be on the safe
side; for such a heavy mediator on--shell production of the mediator
should clearly be negligible, and the EFT limit should be applicable to 
analyses of LHC data. We saw above that for fixed $\Lambda$ the mediator's
width grows like the third power of its mass. It is thus important to 
find the {\em minimal} mass of the mediator for which the predictions
of the simplified model can be reproduced accurately by the EFT.

In order to determine this we again only consider contributions with
$N_{NP}=2$. We compute the monojet cross section after cuts for two
values of the width of the mediator, $\Gamma=1$~GeV and
$\Gamma = M/2$. In case of an axial vector mediator, contributions
with initial $b$ or $t$ quarks are very small, due to their small
parton densities in the proton. For pseudoscalar mediator,
eq.(\ref{eq6}) implies $g = g_{\chi P} \simeq 62$ for $\Lambda = 40$
GeV (near the current bound) already for $M_P = 1$~TeV.
Eq.(\ref{eq4a}) would thus imply a coupling $g_{P t}$ to the top quark
well beyond $\sqrt{4 \pi}$. For only slightly heavier mediator, its
coupling to $b$ quarks would become nonperturbative as well. We
therefore set the couplings to $b$ and $t$ quarks to zero in both
scenarios.  The couplings of the axial vector mediator to the
remaining four quarks are set equal to each other, while the couplings
of the pseudoscalar mediator to these quarks are proportional to the
respective Yukawa couplings, see eq.(\ref{eq4a}).

We then compute the cross section for $\chi \bar{\chi} +$ jet(s)
``monojet'' events after cuts as a function of the mediator mass,
keeping $\Lambda$ fixed. For $\Gamma = M/2$, we include the width
dependence of $\Gamma$, i.e. we replace $M$ by
$\left( M^4 + M^2 \Gamma^2 \right)^{1/4}$ in eqs.(\ref{eq3}) and
(\ref{eq6}). We fix $\Lambda$ to $900 \ (40)$ GeV for axial vector
(pseudoscalar) mediator, close to the current experimental
limits. Since we only include contributions where a single mediator is
exchanged and fix the widths of the mediators, the matrix element is
always proportional to the product of couplings of the mediator to
quarks and to WIMPs. This is true by construction in the EFT limit,
but holds here even for on--shell exchange of the mediator.

\begin{figure}
\centerline{
\includegraphics[scale=0.5]{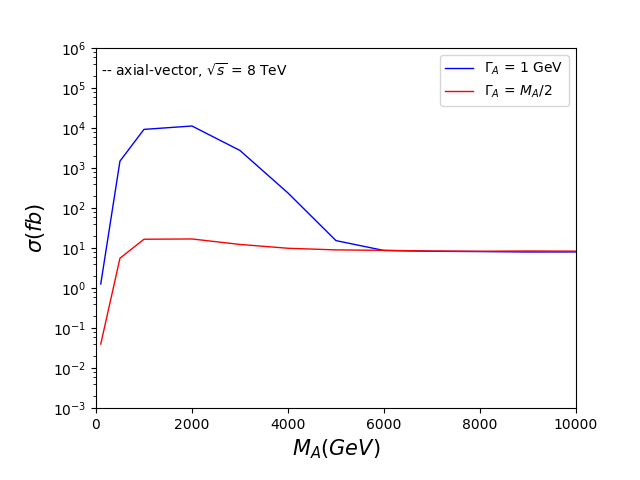}
\includegraphics[scale=0.5]{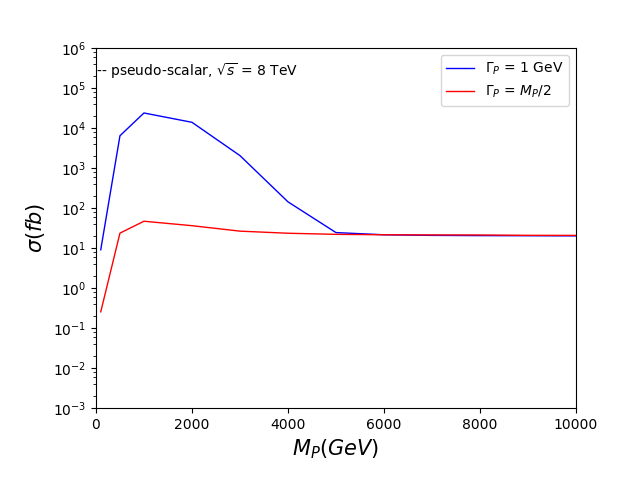}}
\caption{Monojet cross section after cuts for two mediator widths. The
  left frame is for the axial vector mediator with $\Lambda = 900$
  GeV with cuts taken from ATLAS SR7, the right frame for the
  pseudoscalar mediator with $\Lambda = 40$ GeV with cuts taken
  from ATLAS SR6. The couplings have been varied along with the
  mediator masses, such that the scales $\Lambda$ are kept fixed.}
\label{fig:PS_AV_med_diffwidth}
\end{figure}

The results are shown in figs.~\ref{fig:PS_AV_med_diffwidth}. We see that
if we fix the mediator's width to $1$~GeV, as we did in our previous analyses,
it should have a mass of at least $6$~TeV for on--shell production of the
mediator to become negligible. Only for masses above this value does the
cross section become independent of the mediator's mass for fixed $\Lambda$,
as predicted by the EFT. This lower limit is basically the same for axial
vector and pseudoscalar mediator.\footnote{These figures use the ATLAS cuts
that offer the best expected sensitivity in the given model; this differs
slightly, with the AV model favoring a slightly stronger cut on the missing
$E_T$.}

However, taking such a small width exaggerates the problem. Since the
width in this calculation is kept fixed, independent of the mass and
couplings of the mediator, the cross section for on--shell production
of the mediator scales like $1/\Gamma$ after integrating over the
Breit--Wigner peak. An artificially small width therefore implies an
artificially large on--shell cross section. On the other hand,
figs.~\ref{fig:PS_AV_med_diffwidth} also show that even for $\Gamma = M/2$,
at the border of the perturbatively treatable domain, the cross section
becomes approximately independent of the mass only for $M \geq 3$ TeV.

For fixed product $g_A^2 \equiv g_{\chi A} g_{q A}$ the total decay
width of the axial vector mediator is minimized if
$g_{q A}^2 = g^2_A / (2 \sqrt{3})$, giving
\beq \label{gamma_minA}
\Gamma_{A,{\rm min}} = \frac{M_A g_A^2} {\sqrt{3} \pi} 
= \frac{M_A^3} {\sqrt{3} \pi \Lambda^2}\,;
\eeq
here we again assumed equal couplings $g_{q A}$ to all first and
second generation quarks, and have used eq.(\ref{eq3}). Requiring
$\Gamma_A < 0.5 M_A$ for a weakly--coupled theory, and $M_A > 3$ TeV
so that monojet production at the $8$ TeV LHC can be described
adequately by the EFT, thus implies $\Lambda > 1.8$ TeV. This is
``only'' about a factor of $2$ above the lower bound from the $8$ TeV
data. Recall, however, that the signal cross section scales like
$\Lambda^{-4}$; improving the bound by a factor of two would thus
require a $16$ times stronger upper bound on the signal cross section!
We conclude that for parameter choices that give monojet cross
sections near the upper bound, the model with axial vector mediator
{\em cannot} be accurately described by an EFT, if the theory is
weakly coupled, i.e. if perturbation theory is applicable.

\begin{figure}
\begin{center}
\includegraphics[scale=0.55]{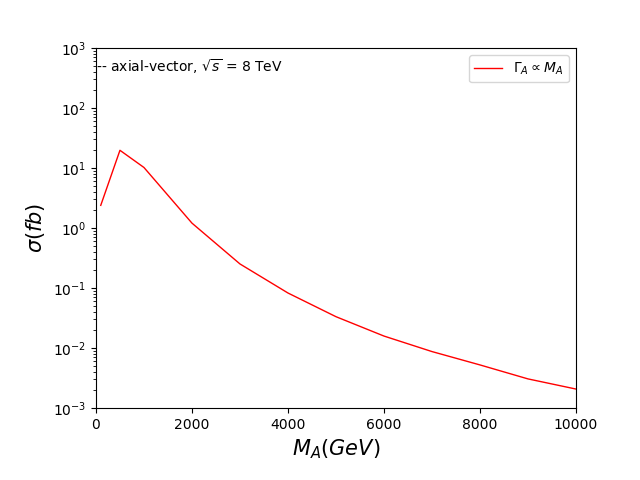}
\caption{Monojet cross section after cuts from ATLAS SR7 in the model
  with axial vector mediator if the mediator's width is calculated
  from eq.(\ref{eq13}). Here the couplings are varied along with $M_A$
  such that $\Lambda$ calculated using eq.(\ref{eq3}) is kept
  fixed at $900$~GeV.}
\label{fig:AV_med_calcwidth}
\end{center}
\end{figure}

This problem can also be illustrated using
Fig.~\ref{fig:AV_med_calcwidth}. This shows the monojet cross section
after the same cuts, and for the same value of $\Lambda$,
calculated from eq.(\ref{eq3}). The main difference is that $\Gamma_A$
has now been computed from eq.(\ref{eq13}), assuming
$g_{\chi A} = g_{q A}$. We see that now there is no region of mediator
mass where the cross section becomes independent of the mass, as one
would expect in the EFT picture. The reason is that, as stated above,
in this case the width grows like the third power of the mass. For the
given choice of couplings, $\Gamma_A > M_A$ for $M_A > 1.5$~TeV.  For
larger values of $M_A$, the cross section drops approximately like
$M_A^{-4}$. Of course, in this regime the theory is no longer weakly
coupled, so the result is not reliable quantitatively.  Note also that
the situation would have been different had $\Lambda$ been an order of
magnitude larger. In this case the cross section would indeed become
almost independent of $M_A$ for some range of masses above $3$~TeV; it
would also be very small, several orders of magnitude below the
experimental bound.

We saw above that already the early (2015) $13$ TeV data slightly
strengthened the bound on $\Lambda$. However, going to higher
center--of--mass energy also requires higher values of $M_A$ for
on--shell production of the mediator to be negligible, so that the
theory can be approximated by an EFT for LHC purposes. For $M_A > 5$
TeV, $\Gamma_A < M_A/2$ is possible only if $\Lambda > 3$ TeV. It seems
extremely unlikely that the upper bounds on the monojet cross section
at the LHC will ever become this strong.

The case of pseudoscalar mediator is slightly different, although the
conclusion will be similar. For the assumed proportionality of the
coupling to a given quark to the mass of this quark, the total decay
width of the mediator is dominated by decay into $c$ quarks and
WIMPs. For fixed product $g_P^2 \equiv g_{\chi P} g$ the total decay
width is minimized if $g^2_{\chi P} = \sqrt{3} m_c g_P^2/v$, giving
\beq \label{gamma_minP}
\Gamma_{P,{\rm min}} = \frac {\sqrt{3} M_P^3 m_c} {4 \pi \Lambda^3}\,.
\eeq
Here we have again assumed that the WIMP is much lighter than the
mediator, and used eq.(\ref{eq6}). We again need $M_P > 3$ TeV for the
EFT to be applicable even if $\Gamma_P = M_P/2$, see
Fig.~\ref{fig:PS_AV_med_diffwidth}. Using a running charm quark mass
$m_c(M_P) = 0.6$ GeV, we find that $M_P > 3$ TeV and
$\Gamma_P < M_P/2$ requires $\Lambda > 110$ GeV. Recall that in
this case the cross section scales like $\Lambda^{-6}$. Reducing
the upper bound on $\Lambda$ from about $40$ to $110$ GeV would
thus require a reduction of the upper bound on the cross section by a
factor of more than $400$. As in case of the axial vector mediator,
the situation is not likely to improve very much at the 13 TeV LHC.

Although we did not treat them explicitly, the cases with vector and
scalar $s-$channel mediators are very similar to those with axial
vector and pseudoscalar mediator, as far as LHC physics is concerned
(although the direct detection limits are much stronger for these
cases, as noted earlier). We are thus forced to conclude that there
is {\em no} weakly coupled simplified model with $s-$channel mediator
to which the monojet bounds derived in the EFT can be applied.

\subsection{Other Constraints}

The parameters of the simplified models are also constrained by considerations
not related to WIMP physics. We will be brief in this Subsection, since we
do not have much new to add to this discussion.

To begin with, a spin$-1$ particle can be described in a
renormalizable theory only as a gauge boson. An axial vector mediator
couples differently to the right-- and left--handed components of
fermions. This can lead to anomalies. Indeed, choosing equal couplings
to all quarks, and to the WIMP, gives non--vanishing anomalies
\cite{Kahlhoefer:2015bea}. These can be canceled by introducing
additional fermions. However, as also pointed out in
\cite{Kahlhoefer:2015bea}, these fermions cannot be much heavier than
the mediator; otherwise unitarity will be violated. If these fermions
can be produced in decays of on--shell mediators, its width will be
even larger than assumed in the discussion of the previous subsection.
Also, some LHC searches for new particles will likely lead to
constraints on the masses of the mediator and the new fermions in this
case. Note finally that these new fermions do not decouple, i.e. their
effect might be felt also at low energies \cite{Dror:2017nsg}.

The mediator can certainly decay into $q \bar q$ pairs. This gives
rise to a bump in the di--jet invariant mass distribution. The fact
that no such bump has been found leads to additional constraints on
the model \cite{Fairbairn:2016iuf}.

Finally, even if the EFT is applicable, models with $s-$channel
mediators inevitably also lead to contact interactions between
four quarks \cite{Dreiner:2013vla}. These have been searched for at the LHC
(and earlier hadron colliders). These searches are usually interpreted
in terms of quark compositeness. The resulting lower bounds on the
scale $\Lambda$ are several times stronger than the ones that have
been derived from monojet searches. These searches probe even
larger momentum exchanges than monojet searches do. It is thus conceivable
that there are model parameters where monojet searches can be interpreted
as an EFT, whereas searches for ``quark compositeness'' cannot. If we 
take these constraints at face value, we would have to assume that the
mediator couples much more strongly to WIMPs than to quarks; note that
the coefficient of the four quark interaction operator is proportional to
the square of the mediator's coupling to quarks, and does not depend on
its coupling to WIMPs.

\section{Inconsistency of the EFT Description}
\label{subsec:results} 

In this Section we show that there is a purely internal inconsistency
with the EFT description as used in the interpretation of monojet
limits, both in the discussion of the previous Sections and by the
experimental groups. The point is that we have so far artificially
restricted ourselves to only including diagrams where a single
mediator is exchanged ($N_{NP}=2$ in the language of MadGraph). Since
we need large mediator masses, and correspondingly large couplings,
for the EFT to describe these contributions accurately, one may wonder
whether diagrams where {\em two} mediators are exchanged ($N_{NP}=4$)
might not be relevant. Here we discuss this for the axial vector
mediator; similar conclusions can be derived for pseudoscalar
mediator.

If the second mediator again couples to a $\bar \chi \chi$ current, we
still need to produce (at least) one hard parton from QCD
interactions. However, if the second mediator couples to a quark
current, {\em no} QCD vertex is required to produce a contribution to
the signal. Note that these contributions always have two partons (quarks
or antiquarks) in the final state. However, we saw above that both ATLAS
and CMS searches tolerate the existence of a second jet in their ``monojet''
searches, if certain conditions are met.

Note that this second mediator can be emitted off a quark line, giving
rise to a matrix element of order $g_{\chi A} g^3_{q A}$, or off the
WIMP line, giving rise to a matrix element of order
$g_{\chi A}^2 g_{q A}^2$. While the latter product is fixed once we
fix the product $g_{\chi A} g_{q A}$ which appears in the matrix
elements with single mediator exchange, the former product depends on
the different combination of these couplings. Hence these
$N_{NP}=4, \ {\cal O}(\Lambda^{-4})$ contributions to the total
matrix element for the signal also depend on the ratio of the two
couplings, in addition to their product.

All $N_{NP}=4$ contributions involve the propagator of one light
fermion (quark or WIMP), in addition to the two mediator
propagators. These contributions can therefore not be expressed as a
single higher--dimensional operator. In some of these contributions
the momentum flowing through both boson propagators is space--like; in
that sense these are double $t-$channel mediators, rather than
$s-$channel mediators. However, there are also several classes of
contributions where the momentum flowing through both mediator lines
is time--like.

\begin{figure}
\begin{center}
\includegraphics[scale=0.4]{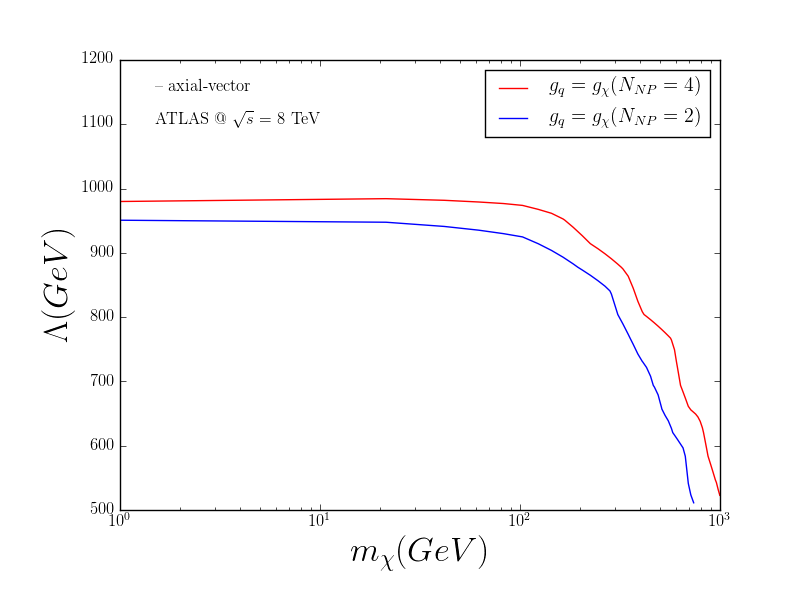}	
\includegraphics[scale=0.4]{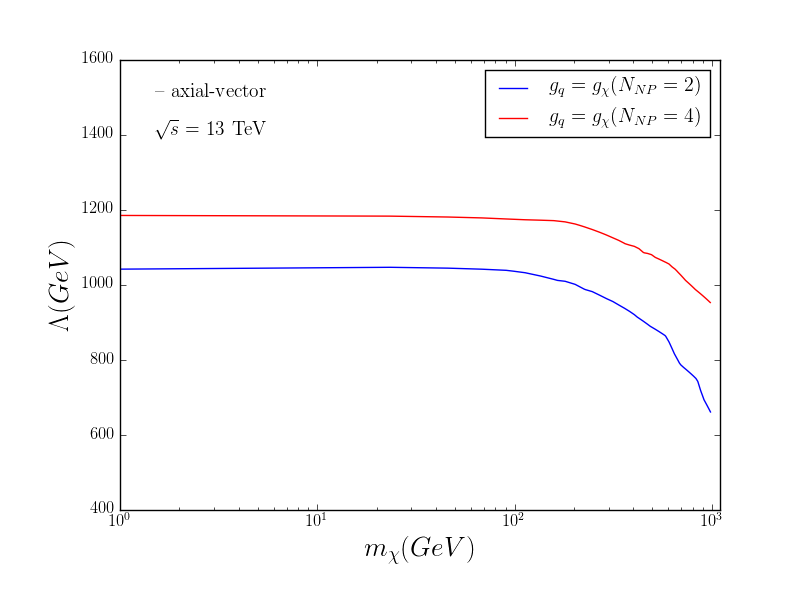}	
\end{center}
\caption{Comparison of bounds on $\Lambda$ at 95\% C.L. derived
  from ATLAS data at $\sqrt{s}=8$~TeV (left) and $13$~TeV (right). The
  blue curves are for $N_{NP} = 2$, i.e. only diagrams where a single
  mediator is exchanged are included, while the red curves also
  include diagrams where two mediators are exchanged ($N_{NP} =
  4$). We have set $g_{\chi A} = g_{q A}$.}
\label{fig:np4vsnp2}
\end{figure}

In Figs.~\ref{fig:np4vsnp2} we compare the bounds on $\Lambda$
that can be derived from ATLAS data for $N_{NP}=2$ and
$N_{NP}=4$. Here we have taken equal couplings of the mediator to all
quarks and to WIMPs. We see that including the $N_{NP}=4$
contributions increases the upper bound on $\Lambda$ by at least
$40$ ($150$) GeV for the $\sqrt{s} = 8$ ($13$) TeV data. An increase of
$40$~GeV, or about $5\%$, does not sound very dramatic. Recall,
however, that the leading contribution to the signal cross section
scale like $\Lambda^{-4}$. A $5\%$ increase of the bound on
$\Lambda$ therefore corresponds to a $20\%$ increase of the {\em
  total} signal cross section; at $\sqrt{s}=13$~TeV, the total cross
section increases by about $60\%$.

The effect of the $N_{NP}=4$ contributions becomes even more
pronounced when we look at specific initial and final states. From the
above discussion it is clear that diagrams with double mediator
exchange always have two partons in the final state; they thus only
contribute to the di--jet part of the signal cross section, which
contributes about $25\%$ of the total cross section after matching if
only the generator--level cuts are applied. Moreover, $N_{NP}=4$
contributions only exist if all external partons are (anti)quarks,
rather than gluons; after the generator--level cuts, for $N_{NP}=2$
all--quark processes contribute about $15\%$ to the total ``di--jet''
cross section, or about $4\%$ of the total signal cross section. The
much stronger final ATLAS cuts enhance the importance of some of these
contributions. In particular, contributions of the kind
$q q \rightarrow \bar \chi \chi q q$ are the only ones with two
valence quarks in the initial state; these contributions suffer the
smallest reduction of the parton densities when the energy scale of
the process is increased by increasing the $\met$ cut. For this
particular class of initial and final states the effect of the
$N_{NP}=4$ contributions is very dramatic. For example, for
$\Lambda = 900$~GeV and $m_\chi = 1$~GeV, the $N_{NP}=4$ terms
increase the cross section for $uu \rightarrow uu \bar\chi \chi$ by a
factor of $2.7$ even if only the generator--level cuts are applied;
here $u$ stands for a $u$ quark or antiquark. The impact of the
$N_{NP}=4$ contributions is even larger after the final cuts, since
the $\Lambda^{-4}$ suppression of these matrix elements implies
that they become relatively more important when the energy scale of
the process is increased.

The relative importance of the $N_{NP}=4$ contributions obviously
increases with decreasing $\Lambda$. Since the bound on
$\Lambda$ decreases with increasing WIMP mass, the impact of the
$N_{NP}=4$ contribution on the final bound is therefore even stronger
for heavier WIMPs.

We thus conclude that for some important subprocesses contributing to
the ``monojet'' signal including only the
${\cal O}(\Lambda^{-2}_{AV})$ contributions to the matrix elements
underestimates the true contribution computed in our simplified model
by a factor $\geq 2$. This is true even in the EFT limit, where
on--shell production of the mediator is negligible.

The importance of these ${\cal O}(\Lambda^{-4})$ contributions to the
matrix element of the signal basically dooms the EFT description. {\em
  Within our simplified model} these contributions can be computed
explicitly. However, in the spirit of an EFT one would for consistency
have to add {\em all} contributions of order $\Lambda^{-4}$. This
includes in particular contributions from operators of mass dimension
up to $8$ in the effective Lagrangian, of which there are a great
many. The usual treatment of ignoring all $N_{NP}=4$ contributions
amounts to the assumption that the coefficients of all of these
dimension$-8$ terms are negligible, which cannot be justified from the
point of view of the EFT alone. Note that this is true already for the
$\sqrt{s}=8$ TeV data. We saw above that, not surprisingly, the effect
of the $N_{NP}=4$ terms is even larger for the $13$ TeV data.

As a purely practical matter, the fact that $N_{NP}=4$ contributions
cannot be neglected means that even in the EFT limit (i.e. for large
mediator mass with artificially small mediator width, and fixed
$\Lambda$) the effective theory has nearly as many relevant
parameters as the simplified model. In both cases the WIMP mass
$m_\chi$ is obviously a free parameter.  In the simplified model, one
in addition needs to fix the mass of the mediator and its couplings to
quarks and to WIMPs; in a true model, its width can then be calculated
from eq.(\ref{eq13}). In the original EFT, these three parameters combine
into the single parameter $\Lambda$ of eq.(\ref{eq3}). However,
as already noted, the $N_{NP}=4$ contributions {\em in addition}
depend on the ratio $g_{\chi A} / g_{q A}$.

\begin{figure}
\begin{center}
\includegraphics[scale=0.55]{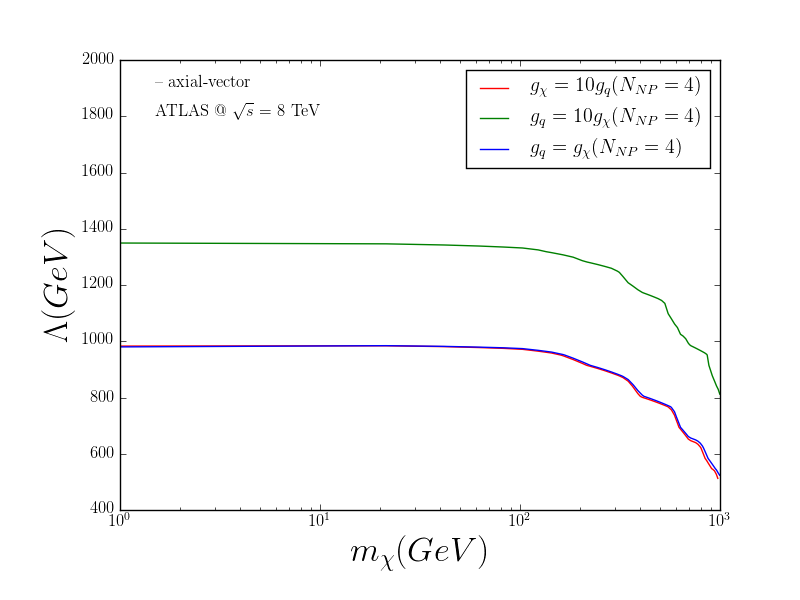}	
\caption{Comparison of ATLAS bounds on $\Lambda$ at 95\% C.L. for
  $N_{NP} = 4$ AV interaction for different coupling ratios
  $g_{\chi A} / g_{q A}$.}
\label{fig:np4gqvsgchi}
\end{center}
\end{figure}

This is illustrated in Fig.~\ref{fig:np4gqvsgchi}. We see that the
effect of the $N_{NP}=4$ contributions becomes {\em much} more
dramatic if the mediator couples more strongly to quarks than to
WIMPs. For example, for $\Lambda = 900$~GeV and
$g_{q A} = 10 g_{\chi A}$, the $N_{NP}=4$ contributions increase the
cross section for $uu \rightarrow u u \bar \chi \chi$ by a factor of
$120$ already after the generator--level cuts; note that if both
quarks in the initial state are valence quarks (rather than
antiquarks), this matrix element is of order
$g_{\chi A} g^3_{qA} = \left( g_{\chi A} g_{q A} \right)^2 (g_{q A} /
g_{\chi A})$,
where the first factor is fixed by $M_A$ and $\Lambda$. Recall, that
scenarios where the mediator couples more strongly to quarks are
constrained very strongly by analyses of di--jet data, as discussed at
the end of the previous Section. However, even if the mediator couples
ten times more strongly to WIMPs than to quarks, the $N_{NP}=4$
contributions increase the cross section for
$u u \rightarrow u u \bar\chi \chi$ by about a factor of two already
after the generator--level cut. Recall that $u$ here denotes either a
quark or an antiquark. The main $N_{NP}=4$ contribution in this case
comes from diagrams where an initial $u \bar u$ pair annihilates into
a virtual mediator, which splits into a $\bar \chi \chi$ pair; the
second mediator is then emitted off the WIMP line, and splits into a
$u \bar u$ pair. This class of diagrams, which scales like
$\left( g_{\chi A} g_{q A} \right)^2$, explains why the total impact
of the $N_{NP}=4$ contributions on the bound on $\Lambda$ becomes
approximately independent of the coupling ratio for
$g_{\chi A} \geq g_{qA}$.

\section{Summary}

In this paper we made the following points:
\begin{itemize}

\item We showed that combining the $\sqrt{s}=8$~TeV data from the
  ATLAS and CMS collaborations increases the bound on the scale
  $\Lambda$ in an effective field theory (EFT) treatment of monojet
  production by a few percent.

\item We showed that there is {\em no} weakly coupled simplified model
  with $s-$channel mediator that can be accurately described by an
  EFT, as far as the LHC monojet analyses are concerned. Here our
  condition for weak coupling was the requirement that the width of
  the mediator is less than half of its mass. The problem arises
  because on--shell production of the mediator becomes negligible only
  for masses that are considerably larger than the scale $\Lambda$ in
  the EFT. This requires large couplings, which in turn leads to large
  decay widths. This problem is even more acute for the pseudoscalar
  mediator.

\item Finally, we showed, that even in an artificial EFT limit (large
  mediator mass with small width), the common practice of only
  including diagrams where a single mediator is exchanged ($N_{NP}=2$
  in the language of MadGraph) underestimates the contributions of
  some subprocesses to the final signal by a factor of more than two
  even if only the generator--level cuts are imposed; the factor
  becomes larger for the final cuts defining the search regions that
  set the final limit. This shows that some contributions to the
  matrix element of the signal that scale like $\Lambda^{-4}$ are
  important. In a consistent EFT one would then have to include {\em
    all} relevant operators with mass dimension up to eight. Such an
  EFT would have many more parameters than a typical simplified model.
  Simply ignoring these terms is, however, not a solution, since some
  such terms {\em always} exist.

\end{itemize}

Our overall conclusion is that the analysis of LHC monojet data in the
language of effective field theory, which was originally hoped to
provide a model independent framework, actually does not apply to {\em
  any} model.

\begin{acknowledgments} 
  SB and MD thank the TR33 ``The Dark Universe'', supported by the
  DFG, for support. SB was also supported by the Deutsche Akademische
  Auslandsdienst (DAAD).

\end{acknowledgments}

\bibliography{eft}
\bibliographystyle{h-physrev} 

\end{document}